\documentclass[pss]{wiley2sp}
\usepackage{amsmath}

\begin{document}
\titlefigure[width=\linewidth,height=3.1cm]{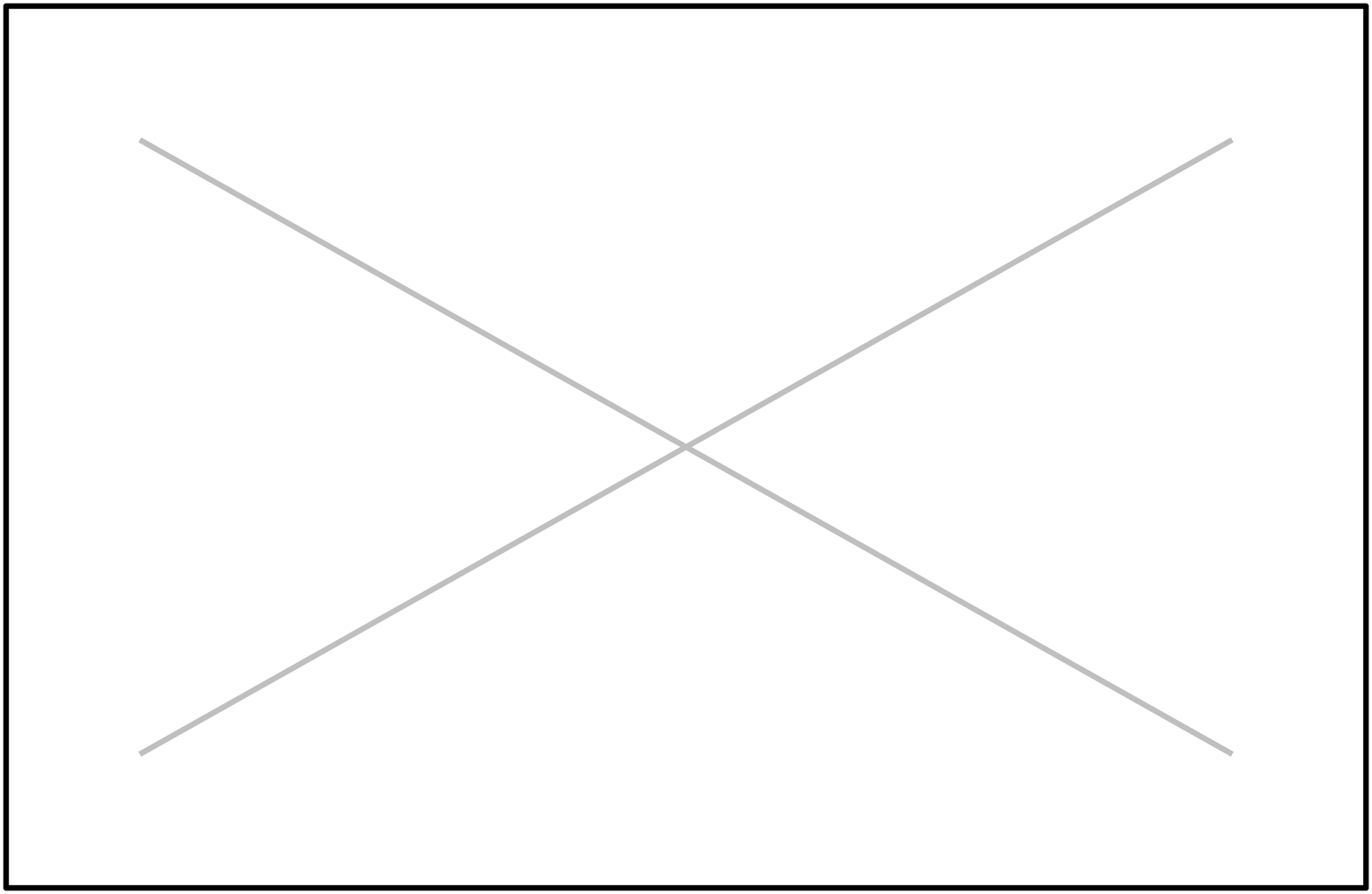}
\titlefigurecaption{This is the figure caption.}
\title{The formation and ordering of local magnetic moments in Fe-Al alloys}
\titlerunning{The formation and ordering of local magnetic moments in Fe-Al alloys}
\author{%
  A.K. Arzhnikov\textsuperscript{\Ast}, 
  L.V. Dobysheva, M.A. Timirgazin}
\authorrunning{A.K. Arzhnikov et al.}
\mail{e-mail
  \textsf{arzhnikov@otf.pti.udm.ru}, Phone
  +07-3412-218988, Fax +07-3412-250614}
\institute{%
  Physical-Technical Institute, 
Ural Branch of Russian Academy of Sciences, 
Kirov str.~132, Izhevsk 426001, Russia}
\received{XXXX, revised XXXX, accepted XXXX}
\published{XXXX}
\pacs{
75.20.Hr; 
75.50.Bb; 
71.15.Ap 
}
\abstract{
With density functional theory, studied are the local magnetic moments
in Fe-Al alloys depending on concentration and Fe nearest environment.
At zero temperature, the system can be in different states:
ferromagnetic, antiferromagnetic and spin-spiral waves (SSW) which
has a minimum energy. Both SSW and negative moment of Fe atoms with many
Al atoms around them agree with experiments. Magnetization curves taken
from literature are analysed. Assumption on percolation character of
size distribution of magnetic clusters describes well the experimental
superparamagnetic behaviour above 150 K.}
\maketitle

\section{Introduction.}

The alloys Fe-Al attracts the attention of researchers as a perspective
material in an extreme technology. They possess the properties such as
good refractoriness, oxidizing and corrosion resistance, relatively low
density, good ductility at room temperature. Intensive study of magnetic
properties was firstly initiated by a development of non-destructive
control methods, as the magnitudes of all the above properties correlate
with magnetic characteristics in these alloys. Afterwards, unusual
behavior of magnetic properties generated a separate  interest to their
study. Mainly the attention is focused on the concentration range from
25 to 50 at.\% of Al for quasiordered and from 40 to 60 at.\% of Al for
disordered alloys. Reliably enough, it was established that in higher
and lower concentration ranges the alloys are in the ferromagnetic and
paramagnetic states, correspondingly, and in the intermediate region
more complicated states are realized. To the beginning of the 80-s a
majority of researchers have been convinced that at low temperature the
magnetic state is a spin glass in this region. But a row of experimental
data that had been then considered as an evidence for the spin glass,
had, as a matter of fact, another nature, which has been revealed later.
For example, the thermomagnetic hysteresis in these alloys is a
consequence of a solely magnetic hysteresis \cite{Els3}. Recently, the
neutron powder diffraction has shown that at low temperature the
magnetic order is governed by spin-density waves \cite{NoakesPRL03}. At
high temperature the alloys are superparamagnetic \cite{Els3,Voronina}.

A raw of discrepancies in studies of magnetic properties of these alloys
has given an impetus to our paper. Here, a theoretical analysis of the
superparamagnetic behavior in the experimental data available is
conducted, collinear and spiral magnetic structures are studied with
the help of the density-functional theory.

\section{Superparamagnetic behavior.}

Magnetization curves of the alloy with 34 at.\% Al as a function of the
applied magnetic field and temperature have been received in the
Ref.\cite{Voronina}. 
\begin{figure*}[!tbh]
\includegraphics[width=6.5in]{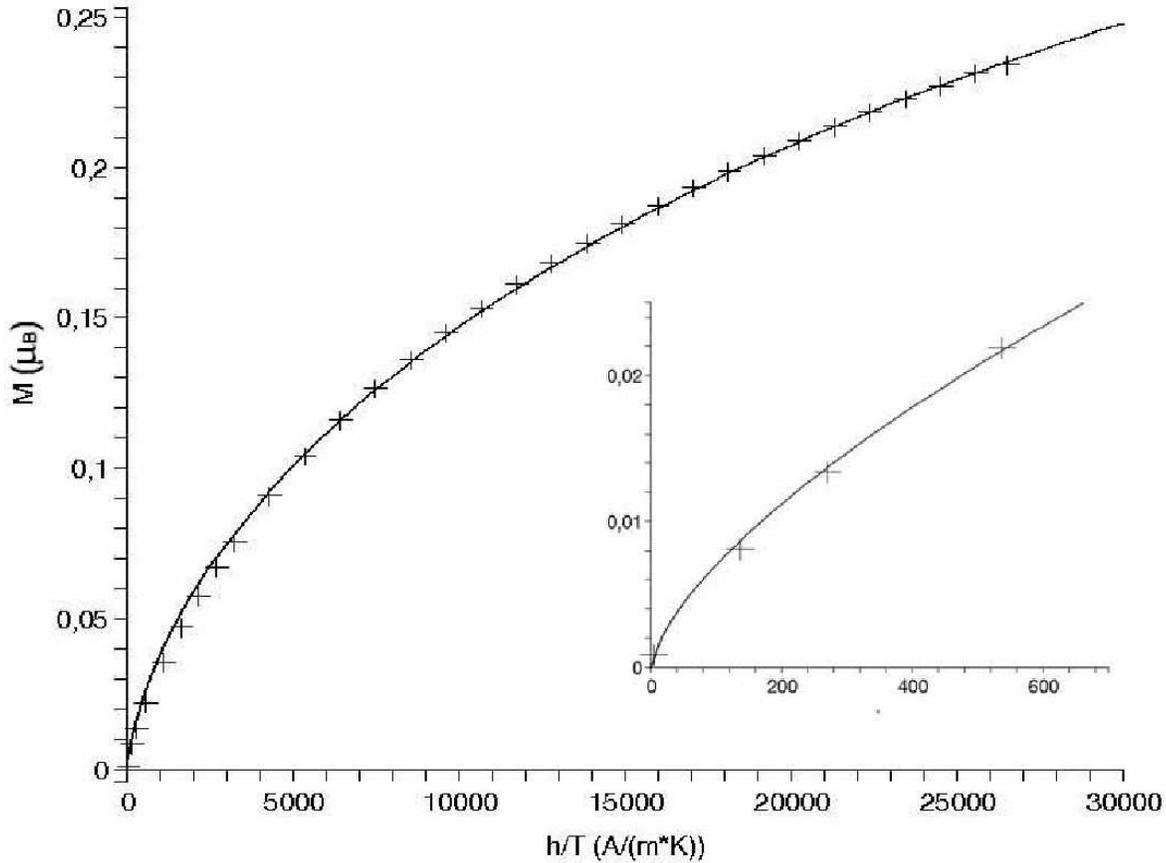} 
\caption{The magnetization curves fitting the experimental data from 
Ref.\cite{Voronina} in assumption about a continuous cluster size
distribution.}
\label{Mfin}\end{figure*}
Higher than the blocking temperature of 150 K the magnetization curves
join each other, which is typical for the paramagnetic behavior.
Besides, the magnetization increases rather quickly at low magnitude of
the parameter $h/T \approx 10^3 A/m/K$, but does not reach the
saturation up to $5 \times 10^4 A/m/K$, which is an evidence for
different by size magnetic clusters that do not interact. Note that the
system is structurally homogeneous, so the clusters are governed by
magnetic interactions.  Using the Arrhenius formula for the relaxation
time $t$ one can estimate the upper limit for the number of atoms in
clusters. It is an order of $n_{max} \approx 10^4$ atoms (the 
characteristic time of magnetic measurement at which the detection 
of largest clusters is possible equals $t=t_0 \times
exp\{-n_{max}E_a/KT\}\approx 10^{-2}$ s, where $E_a \approx 7 \times 10^{-25}
J/at$ is energy of the magnetocrystalline anisotropy in iron,
$t_0\approx 10^{-6}$ c is the spin-lattice relaxation time, $T=150 K$ is
a blocking temperature).

Assuming that the alloy contains clusters of two types, we succeeded in
describing the magnetization curves with clusters of 6 nm diameter (6500
atoms in the cluster) and 3 nm diameter (600 atoms in the cluster). 20
\% of all atoms belong to the 6 nm clusters and 80 \% of all atoms
belong to the 3 nm clusters. The average magnetic moment of an Fe atom
is $0.33 \mu_B$. This description has however essential shortcomings.
First, there is no physical or chemical mechanism which could be
responsible for just these cluster sizes (as authors of
Ref.\cite{Voronina} assure, the sample was homogeneous). Second, with
such size distribution of clusters, magnetization at weak fields should
be proportional to h/T, which is not corroborated by experiment (see
insert in Fig.~\ref{Mfin}). 

More naturally looks the assumption about a continuous size distribution
as this is in the case of hierarchy of the cluster size distribution in
disordered percolation task \cite{Kirkp,Isich}. In this case, density of 
the number of clusters consisting of $n$ Fe atoms divided by total
number of lattice sites is equal to  $w_n = x(\tau -2) n_{min}^{\tau-2}
n^{-\tau}$, where $\tau$ is a critical exponent \cite{Kirkp,Isich},
$n_{min}$ is a minimal cluster size and $x$ is concentration of Fe
atoms. As the magnetic moment of a cluster is large, one can use
classical concepts and calculate the magnetization per iron atom (see
Ref.\cite{White}):
\[\begin{array}{l}
M=m_{av}(\tau -2)n_{min}^{\tau -2} \times \\  \; \; \; \; \; \; \; \; \; \; \; \; \; \;  
\int^{\infty}_{n_{min}}  n^{1- \tau }[cth(n m_{av} h/kT)-kT/n m_{av} h]\;dn
\end{array} \]
Here $m_{av}$ is average magnetic moment of an Fe atom in a cluster.

Scaling relations \cite{Isich} allow us to write $\tau=\delta^{-1}+2$,
where $\delta$ determines the magnetization behavior $M \propto
(h/T)^{1/\delta}$ at $h/T \to 0$. In the insert to Fig.~\ref{Mfin},
shown is a least-squares adjustment of the experimental data with
$\delta = 1.49$.  From the above interrelation between coefficients we
receive $\tau = 2.67$. Further, using this $\tau$, we have conducted
fitting in the whole range of the parameter $h/T$ (Fig.~\ref{Mfin}). The
best agreement with experiment have been achieved at $n_{min}=62$ atoms
and $m_{av}=0.44\mu_B$. 

We must note that the coefficient $\tau$ obtained does not coincide by
magnitude with the coefficient in the classical percolation theory
($\tau=2.2$ \cite{Isich}). To our opinion, the reasons are the
following: first, the sample in the experiment was quasiordered, that
is, had a strong short order; second, it did not reach the percolation
threshold;  third, the interactions between the atomic magnetic moments,
that govern the geometry of magnetic clusters, are connected with the
chemical configurations of the atoms disposition in a very complicated
way.

\section{Dependence of Fe magnetic moments on the closest atomic
environment.}

To understand the peculiarities of the magnetic interaction in the Fe-Al
alloys we have conducted first-principles calculations of the periodical
systems $Fe_{38}Al_{16}$ (29.6 at.\% Al), $Fe_{11}Al_5$ (31.3),
$Fe_{34}Al_{20}$ (37.0), $Fe_{10}Al_6$ (37.5) and $Fe_9Al_7$ (43.8).
These systems have been chosen so that to cover that interesting
intermidiate concentration region from 29 to 44 at.\% Al and to receive
many different chemical configurations of the iron nearest environment.
The calculations have been conducted by FP LAPW method with the WIEN 2k
program package \cite{WIEN2k}. The detailed description of the models
and approximations used is given in \cite{ArzhJETF2007}.

\begin{figure}[!ht]  
\includegraphics[width=3.5in]{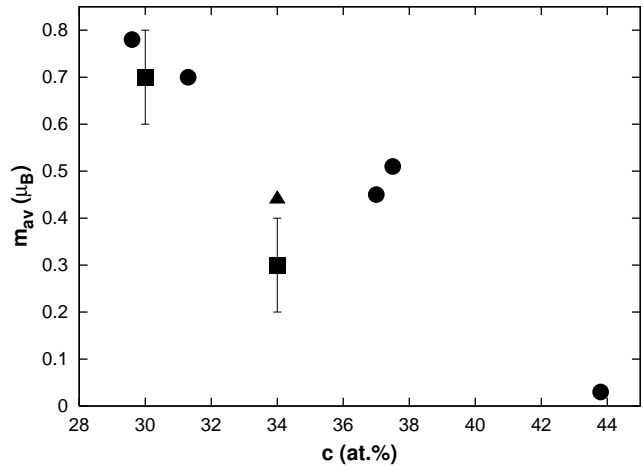} 
\caption{Average magnetic moment of Fe atoms as a function of Al
concentration. Circles denote the first-principles results, squares are
for experimental data from Ref.\cite{Els4}. Triangle shows the
moment obtained from analysis of superparamagnetic behavior.}
\label{M_aver_c}\end{figure}

\begin{figure}[!ht] 
\includegraphics[width=3.5in]{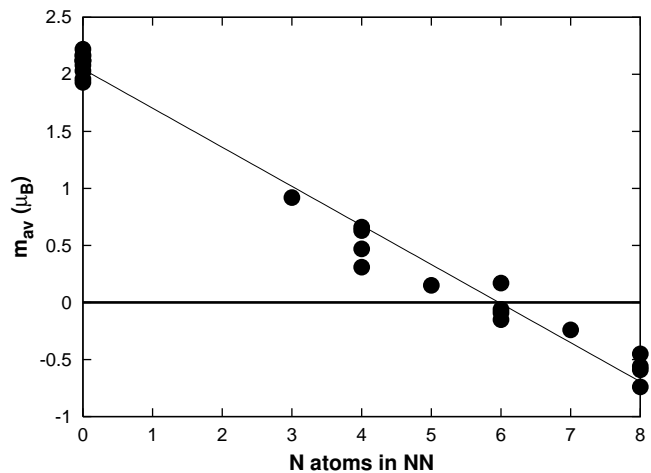} 
\caption{The local magnetic moment of Fe atoms as a function of number
of Al atoms in the nearest environment.}
\label{M_N}\end{figure}

One of the main results consists in the following: there are two
solutions with collinear magnetic moments found for all the
concentrations studied. One of them has Fe local magnetic moments all of
a direction, the other has both positive and negative moments depending
on environment. Namely, the magnetic moments at Fe atoms with 6 and more
Al atoms in nearest environment direct oppositely to those of the rest
iron atoms. In the following, we call the first as the solution of a
ferromagnet type (FM), and the second as that of an antiferromagnet type
(AFM). The AFM states are slightly lower by energy than the FM ones in
$Fe_{11}Al_5$, $Fe_{34}Al_{20}$ and $Fe_{10}Al_6$; in $Fe_{38}Al_{16}$
and $Fe_9Al_7$ the FM state is more preferable. The Fe average magnetic
moment in the AFM solutions as a function of Al concentration agrees
rather well with experimental data (Fig.~\ref{M_aver_c}).
Fig.~\ref{M_N} shows the iron magnetic moments in all the systems
studied, in the aggregate, depending on nearest environment. 
One can see that direction and magnitude of the magnetic moments in AFM
solutions are rather accurately determined by chemical composition of
the iron nearest environment, and only small variations can be imputed
to different structure, concentration or environment in more distant
coordination spheres. 

Similar results have been also obtained earlier for disordered alloys
using a two-band Hubbard model \cite{ArzhJMMM1992}. This behavior of
local magnetic moments gives grounds for usage of the Jaccarino-Walker
model for interpretation of experiments. The main idea of the Jaccarino-
Walker model asserts that the local magnetic moment at a
transition-metal atom is determined by chemical composition of the
nearest environment and only weakly depends on the overall
concentration. Using this model authors of Refs.\cite{Els4,Els2} have
described combination of magnetic and Mossbauer experimental data in
disordered and partly disordered Fe-Al alloys. Surely, the models
describing the magnetic order in terms of closest environment cannot
pretend to be very precise in the transition-metal alloys. Nevertheless,
this model considering the closest environment is much better in
comparison with commonly used models of the Heisenberg type where the
magnetic moments of closest atoms interact ferromagnetically and the
magnetic moments of iron atoms with an aluminium atom between interact
antiferromagnetically \cite{NoakesPRL03,SatoArrott57}.

\section{Spiral magnetic structures.}

Recently, neutron diffraction studies  \cite{NoakesPRL03} have shown
that quasiordered Fe-Al alloys have spin-density waves with [111]
direction at temperature lower than 100 - 150 K. Such study, as authors
of Ref.\cite{NoakesPRL03} themselves admit, cannot distinguish the
spin-density wave with a collinear structure (SDW) and the spin-spiral
wave (SSW). SDW in Cr have been rather long ago known and well studied.
A nesting in the Fermi surface \cite{Overhaus} is considered the most
justified mechanism for appearance of SDW in Cr. This mechanism looks
impossible for the Fe-Al alloys. First, the Fermi surface of iron does
not have nesting; second, the alloys have a disorder in the atomic
disposition, which makes the influence of the Fermi surface on the SDW
formation very problematic. That is why we think that the oscillations
in the experiment come from SSW and consider the conditions of their
appearance in the systems $Fe_9Al_7$ and $Fe_{10}Al_6$. The calculation
is conducted with use of a  non-collinear-magnetic version of WIEN2k
package \cite{NCM1,NCM2}. The SSW in  $Fe_9Al_7$ has been considered earlier
in \cite{Bogner}. They have received that the SSW with [001] direction
and the wave vector $q=0.4 a_0^{-1} $ possesses the minimum energy (here
$a_0$ is a bcc lattice parameter). We have received the same result for
$Fe_9Al_7$. For $Fe_{10}Al_6$ the minimal by energy SSW direction
coincides the experimental one [111]. The difference between the
collinear and the SSW solutions is less than 7 mRy/cell which is a small
value and allows transitions from SSW to a collinear state at small
energy of an external excitation (magnetic field or temperature). We
must note that wavelength of the SSW received in both our and
Ref.~\cite{Bogner} calculations is $4 a_0$, while the experimental value
observed at these concentrations is $7 a_0$ \cite{NoakesPRL03}.

We did not take into account the spin-orbit interaction in our
calculations, so it cannot be responsible for the appearance of the SSW
as this usually occurs in the magnetics on the basis of rare-earth
elements and actinides. To our opinion, the main reason of the SSW
appearance as the ground state is a competition of the two collinear
magnetic states FM and AFM that are close by energy.

\section{Conclusions.}

At temperatures higher than 150 K the alloy with 34 at.\% of Al is a
typical superparamagnetic \cite{Voronina}. The best theoretical
description of experimental magnetization curves is obtained with
assumption that cluster distribution by size obeys a scaling law with
minimum size clusters as $\approx$ 60 magnetic atoms, and with the
average local magnetic moment of Fe atom is $m_{av}=0.44 \mu_B$.

Our study has shown a potential possibility of existence of few types of
magnetic order in Fe-Al alloys: collinear structures (FM and AFM) and
spin-spiral waves (SSW). The energy of SSW is lower than those of the FM
and AFM structures. The difference in energy between these states does
not exceed 7 mRy/cell. This allows the system to transform from one
magnetic structure to another at weak external influence (magnetic field
or temperature). The character of the thermal or field transition from
SSW to a collinear state is, however, unclear in actual disordered
alloys: is it a kind of phase transition or the transition occurs
through a row of continuous reconstructions of the electron structure in
local regions?

The average Fe magnetic moments theoretically calculated in structures
with AFM ordering (Fig.~\ref{M_aver_c}) are close to the experimental
data from direct magnetization measurements and from the analysis of 
the superparamagnetic behavior. This fact allows researchers to use for
interpretation of experimental data a modified Jaccarino-Walker model,
with a dependence of the magnetic moment on the closest environment
similar to Fig.~\ref{M_N}.

\begin{acknowledgement}
This work was partially supported by RFBR (grant 06-02-16179-a).
\end{acknowledgement}

\end{document}